\documentclass[letterpaper, conference]{ieeetran}

\usepackage{amsmath}
\usepackage{caption}

\usepackage{balance}
\usepackage{cite}
\usepackage{float}
\usepackage{multicol,multirow}
\usepackage{makecell,booktabs}
\usepackage{hyperref}
\hypersetup{    
    colorlinks=true,
    linkcolor=black,
    anchorcolor=black,
    citecolor=black,
    filecolor=black,
    menucolor=black,
    runcolor=black,
    urlcolor=black,
}

\usepackage{subcaption}
\usepackage[utf8]{inputenc}

\usepackage{graphicx}
\usepackage{amsfonts}
\usepackage{amssymb}
\usepackage{amsmath,mathtools}
\usepackage{array}
\newcolumntype{P}[1]{>{\centering\arraybackslash}p{#1}}
\newcolumntype{L}[1]{>{\raggedright\arraybackslash}p{#1}}
\newcolumntype{R}[1]{>{\raggedleft\arraybackslash}p{#1}}
\usepackage{color}

\usepackage[labelsep=period]{caption}
\usepackage{bbm}
\usepackage{bm}
\usepackage{comment}
\usepackage{forest}
\usepackage{dblfloatfix}

\usepackage{etoolbox}
\usepackage{kotex}

\usepackage{tabularx}

\title{\bf \Large
    A Risk-aware Bi-level Bidding Strategy for Virtual Power Plant \\
    with Power-to-Hydrogen System
    {\vspace{-4mm}}
}

\author{\IEEEauthorblockN{Jaehyun Yoo and Jip Kim*}
\IEEEauthorblockA{Dept. of Energy Engineering, Korea Institute of Energy Technology\\
\{jaehyuny, jipkim\}@kentech.ac.kr}

\thanks{This work was supported by National Research Foundation of Korea (NRF) grant funded by the Korean government (MIST) (No. RS-2023-00210018) and by the Korea Institute of Energy Technology Evaluation and Planning (KETEP) and the Ministry of Trade, Industry and Energy (MOTIE) of the Republic of Korea (No. RS-2023-00236325) }
}

\begin{document}

\IEEEoverridecommandlockouts

\maketitle

\IEEEpubidadjcol

\begin{abstract}
This paper presents a risk-aware bi-level bidding strategy for Virtual Power Plant (VPP) that integrates Power-to-Hydrogen (P2H) system, addressing the challenges posed by renewable energy variability and market volatility.
By incorporating Conditional Value at Risk (CVaR) within the bi-level optimization framework, the proposed strategy enables VPPs to mitigate financial risks associated with uncertain market conditions. The upper-level problem seeks to maximize revenue through optimal bidding, while the lower-level problem ensures market-clearing compliance.
The integration of the P2H system allows surplus renewable energy to be stored as hydrogen, which is utilized as an energy carrier, thereby increasing market profitability and enhancing resilience against financial risks.
The effectiveness of the proposed strategy is validated through a modified IEEE 14 bus system,
demonstrating that the inclusion of the P2H system and CVaR-based risk aversion enhances both revenue and financial hedging capability under volatile market conditions.
This paper underscores the strategic role of hydrogen storage in VPP operations, contributing to supporting improved profitability and the efficacy of a risk-aware bidding strategy.
\end{abstract}

\section{Introduction}\label{Sec:Intro}
The increasing integration of renewable energy sources (RES) and distributed energy resources (DERs) into modern power grids is reshaping the operational dynamics and strategies required for effective grid management.
This shift has intensified the need for flexible management of energy assets, as the inherently intermittent nature of RES, such as photovoltaic (PV) and wind turbine (WT), requires sophisticated coordination to maintain grid stability.
Virtual Power Plant (VPP) has emerged as a pivotal solution, aggregating diverse DERs—including PV, WT, and energy storage system (ESS)—into a centralized framework that optimizes their participation in markets \cite{VPP_9}.
VPP allows small resources to be managed as a unified entity, enabling them to compete with conventional power plants by capitalizing on their aggregated capacity.
This aggregated structure supports seamless participation in the market, promoting efficient energy distribution in the face of increasing renewable penetration.

Despite these advancements, VPP operators encounter significant challenges stemming from the inherent uncertainty in renewable generation and the volatility of market prices.
Fluctuations in energy output and market volatility introduce substantial risks, particularly when the VPP heavily depends on RES, threatening their economic viability.
To address this challenge, researchers have increasingly focused on bi-level optimization model to enhance the strategic bidding capabilities of VPP \cite{VPP_2, VPP_7, VPP_10}.
Typically, this bi-level model consists of an upper-level problem, where VPP aims to maximize revenue through optimized bidding strategies, and a lower-level problem, which incorporates market-clearing to ensure feasible operations within the competitive market landscape.

However, traditional profit-maximization methods prioritize only the maximization of VPP revenue \cite{VPP_15}, resulting in full exposure to substantial risks associated with high market volatility.
In \cite{VPP_12}, neglecting risk consideration in highly volatile markets resulted in an approximate 10\% revenue loss.
Given the limitations of such conventional approaches, there is a growing emphasis on integrating risk mitigation measures into VPP optimization frameworks.
Conditional Value at Risk (CVaR), for instance, has emerged as a critical tool for managing risks by focusing on the potential extreme losses rather than average scenarios, providing a more robust approach for VPP to withstand adverse market conditions \cite{Book_1, VPP_7, VPP_12}.By incorporating CVaR, VPP operators can prioritize strategies that effectively hedge against the potential financial impacts of severe market fluctuations.

The advent of the Power-to-Hydrogen (P2H) system enhances VPP operation by converting surplus electricity into hydrogen during low demand periods \cite{P2G_1, P2G_2}.
Additionally, projections indicate that by 2030, the levelized cost of storage (LCOS)—a metric assessing storage system cost combining capital and operational expenditures, storage duration, and discharge cycles—of the P2H system will be lower than that of Li-Ion ESS \cite{P2G_10, P2G_9}. 
Despite its potential, research on integrating the P2H system within VPP framework is limited, with few studies exploring its economic impacts under varying market conditions \cite{P2G_3, P2G_4, P2G_6}.
The P2H system functions like ESS in power supply and absorption but enables simultaneous charging and discharging via the hydrogen storage system (HSS), utilizing an electrolyzer and fuel cell.
Yan Meng \textit{et al.} \cite{P2G_3} include electrolyzer production rate and tank pressure via a compressor, but the framework does not encompass fuel cell in HSS. 
Qunli Wu \textit{et al.} \cite{P2G_4} include both an electrolyzer and a fuel cell, yet concurrent participation of both types of equipment was restricted. 
Rodrigues \textit{et al.} apply \cite{P2G_6}, non-constant efficiency curves, but this nonlinear form limits its applicability in bi-level optimization.
Overall, integrating HSS within the VPP framework is still constrained by challenges in the operation of electrolyzer and fuel cell simultaneously and applying bi-level optimization to address nonlinear efficiency in HSS.

\begin{figure}[t]
    \centering
\includegraphics[width=1\columnwidth]{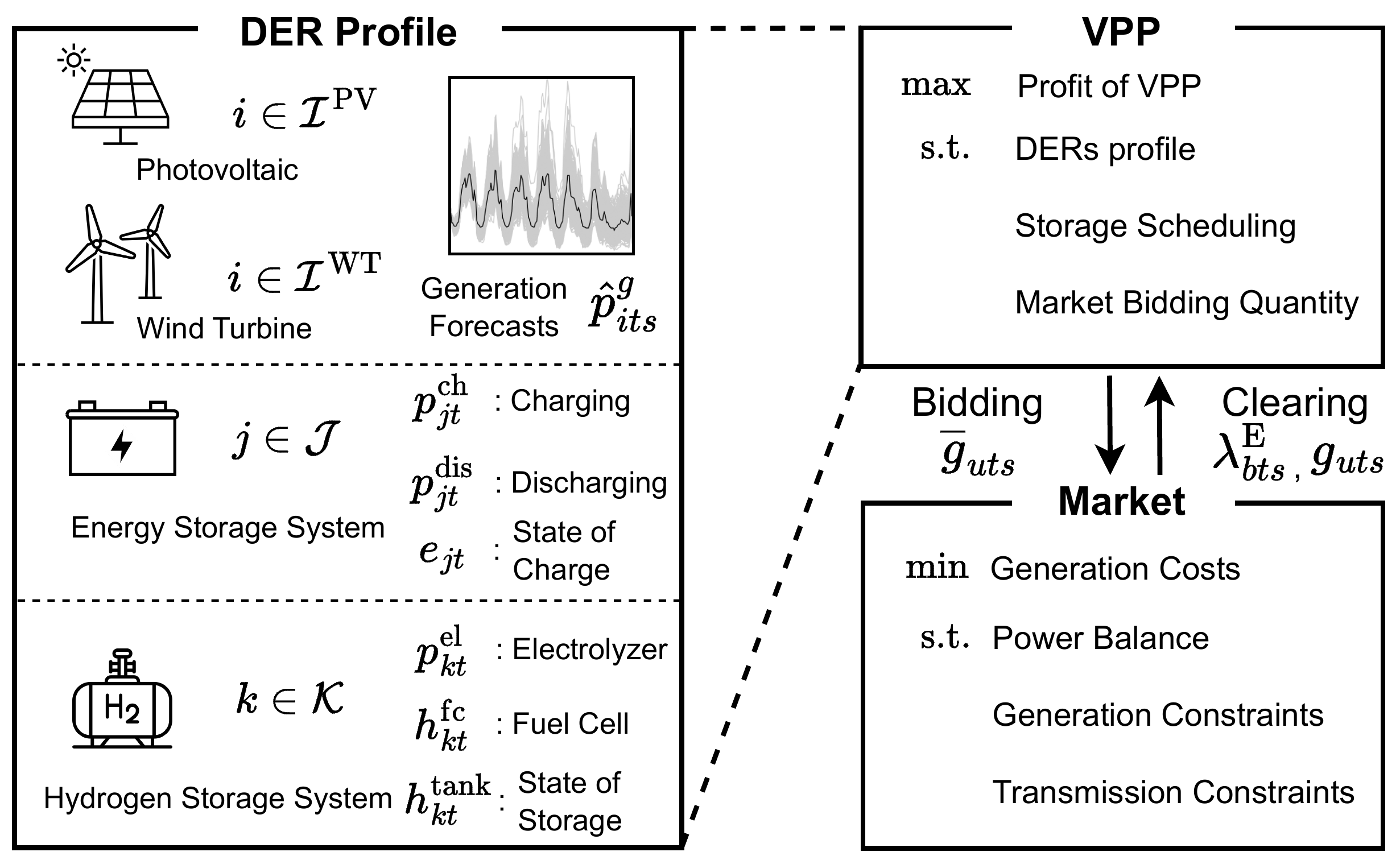}
\vspace{-3mm}\caption{Framework of the Proposed Model}\vspace{-7mm}
\label{fig:VPP framework}
\end{figure}

To overcome the limitations, we propose a risk-aware bi-level bidding strategy that integrates a P2H system within the VPP framework.
The main contributions of this paper are:

\begin{itemize}
\item Application of CVaR: This paper applies CVaR within the VPP optimization framework to address the inherent risks posed by market volatility.
\item Bi-level optimization problem: Bi-level optimization framework is designed to adapt bidding strategies to fluctuating RES output and market prices.
\item Concurrent operation of HSS: By enabling simultaneous operation of the electrolyzer and fuel cell, VPP can enhance the profitability and robustness to risk aversion.
\end{itemize}

\section{Proposed Model}\label{Sec:Formulation}

Figure ~\ref{fig:VPP framework} illustrates the overall framework of the proposed model, outlining the key components of VPP and its interaction with the market. 
In the upper-level problem, based on probabilistic scenario-based forecasts of generation from RES, the VPP optimizes the scheduling of its assets to maximize revenue. 
Based on generation forecasts for PV and WT, the optimal bid scheduling for ESS, and HSS in the VPP is determined. In response to varying conditions, the ESS can supply or absorb power but cannot operate at the same time. 
In contrast, the HSS operates similarly to ESS by supplying or absorbing power, yet, due to the electrolyzer for supply and fuel cell for absorption, it achieves charging and discharging simultaneously.
In the lower-level problem, by adapting its bid quantity based on the status of individual assets, the VPP acts as a single generator in the market. 
The cleared generation quantity and electricity price are influenced by the strategic VPP and competitor bids, as well as anticipated load levels. 
The VPP can consider the price effect of its bids within its bid schedule to maximize revenue.

\vspace{-1mm}
\subsection{Upper-level Problem: VPP Bidding Model}

1) Objective Function: We formulate the objective function as the sum of expected revenue and CVaR of the revenue for VPP, which is modulated by the weighting coefficient $\beta$.
\begin{subequations}
\begin{align}\label{eq:obj}
\max_{\Xi^{\rm{UL}}} \quad & 
    \sum_{s \in {\cal S}}  \sum_{t \in  {\cal T}} \omega_{s}  (
    \lambda^{\rm{E}}_{b(\mathrm{VPP})ts} g^{\rm{VPP}}_{uts} 
    + \sum_{i \in {\cal I }} \kappa^{\mathrm{REC}}_{ts} p^{\rm{g}}_{its}  ) \nonumber \\
    & + \beta \cdot ( \eta - \frac{1}{1-\alpha}\sum_{s \in {\cal S}  } \omega_{s} \epsilon_{s}  )
\end{align}
where  $\Xi^{\rm{UL}} \!\!\coloneqq\!\! \{ p^{\rm{dis}}_{jt}, p^{\rm{ch}}_{jt}, p^{\rm{el}}_{kt}, h^{\rm{fc}}_{kt} \!\in\! \mathbb{R}; \,\, z_{jt} \!\in\! (0,1) \}$.
$\omega_{s}$ denotes the probability of each scenario $s$, $t$ represents the operational time interval, and $i$ is the index for VPP generation units PV and WT.
Here, $\lambda^{\rm{E}}_{bts}$ denotes the market price determined by the lower-level problem for each scenario, with the subscript $b(\rm{VPP})$ indicating the node to which the VPP belongs.
The profit in the market for each scenario is then obtained by multiplying the market price by the awarded generation of VPP $g^{\rm{VPP}}_{uts}$.
Renewable Energy Certificate (REC) profit $\kappa^{\mathrm{REC}}_{t}$ is additionally obtained from the generation output $p^{\rm{g}}_{its}$ of its renewable energy assets PV and WT, which is related to  $\lambda^{\rm{E}}_{bts}$.
It is assumed that all resources are owned by the VPP, and no installation costs are considered.
The description of the remaining term is presented in the description of CVaR constraints 4) explained below.

2) VPP DER Constraints: The VPP, comprising PV, WT, ESS, and HSS units, utilizes probabilistic scenario-based forecasts for maximum PV and WT generation, enabling ESS and HSS to optimize bid schedules aimed at revenue maximization.
\begin{align}\label{eq:DER}
\text{s.t.} \quad 
& \kappa^{\mathrm{REC}}_{ts} = 0.5 \lambda^{\rm{E}}_{b(\mathrm{VPP})ts}, \forall t \in {\cal T}, \forall s \in {\cal S}, \\
& 0 \leq p^{\rm{g}}_{its} \leq \hat{p}^{\rm{g}}_{its}, \quad \forall i \in {\cal I}, \forall t \in {\cal T}, \forall s \in {\cal S}, \\
& 0 \leq p^{\rm{ch}}_{jt} \leq \overline{p}^{\rm{ch}}_{jt} \cdot z_{jt}, \quad  \forall j \in {\cal J}, \forall t \in {\cal T},  \\
& 0 \leq p^{\rm{dis}}_{jt} \leq \overline{p}^{\rm{dis}}_{jt} \cdot (1 - z_{jt}) , \quad \forall j \in {\cal J}, \forall t \in {\cal T},  \\
& e_{jt} = e_{jt-1} + \eta^{\rm{ch}}_{j} p^{\rm{ch}}_{jt} - p^{\rm{dis}}_{jt} / \eta^{\rm{dis}}_{j}, \!\!\! \quad \forall j \in {\cal J}, \forall t \in {\cal T}, \!\!  \\
& e_{jt=0,24} = 0.5 \overline{e}_{j}, \quad \forall j \in {\cal J},   \\
& 0.2 \overline{e}_{j} \leq e_{jt} \leq 0.8 \overline{e}_{j}, \quad \forall j \in {\cal J}, \forall t \in {\cal T},\\
& 0 \leq p^{\rm{el}}_{kt} \leq \overline{p}^{\rm{el}}_{kt}, \quad \forall k \in {\cal K}, \forall  t \in {\cal T},\\
& 0 \leq h^{\rm{fc}}_{kt} \leq \overline{h}^{\rm{fc}}_{kt}, \quad \forall k \in {\cal K}, \forall  t \in {\cal T},\\
& h^{\rm{tank}}_{kt} \!\! =\! \! h^{\rm{tank}}_{k,t-1} \!+\! \eta^{\rm{el}}p^{\rm{el}}_{kt}/M^{\rm{H_{2}}} \!\!-\!\! h^{\rm{fc}}_{kt},  \forall k \in {\cal K}, \forall t \in {\cal T}, \!\!\\
& 0.4 \overline{h}_{k} \leq h^{\rm{tank}}_{k,t=0,24} \leq 0.6 \overline{h}_{k}, \quad  \forall k \in {\cal K},\\
& 0.2 \overline{h}_{k} \leq h^{\rm{tank}}_{kt} \leq 0.8 \overline{h}_{k} , \quad  \forall k \in {\cal K}, \forall  t \in {\cal T},
\end{align}
Equation (1b) represents the relationship between REC price $\kappa^{\mathrm{REC}}_{ts}$ and 
the market price $\lambda^{\rm{E}}_{b(\rm{VPP})ts}$, where $\kappa^{\mathrm{REC}}_{ts}$ is half of $\lambda^{\rm{E}}_{b(\rm{VPP})ts}$.
Equation (1c) describes the maximum generation forecast $\hat{p}^{\rm{g}}_{its}$ for PV and WT based on probabilistic scenarios.
Equations (1d)-(1h) define conditions for the ESS bidding schedule, with binary variable $z_{jt}$ in (1d) and (1e) directing charge operation $p^{\rm{ch}}_{jt}$ and discharge operation $p^{\rm{dis}}_{jt}$.
Additionally, Equations (1f)-(1h) define the operable state of charge (SoC) range $e_{jt}$ for ESS throughout the day, with initial and final SoC values aligned to market start and end conditions.
Equations (1i)-(1m) describe the bid scheduling of the HSS, comprising the electrolyzer, hydrogen storage tank, and fuel cell.
Hydrogen produced by the electrolyzer is stored in the tank and used for electricity generation via the fuel cell, enabling simultaneous charging and discharging.
Equations (1i) and (1j) represent the power consumption of the electrolyzer $p^{\rm{el}}_{kt}$ and the hydrogen usage of the fuel cell $h^{\rm{fc}}_{kt}$, with both devices operating simultaneously.
Equations (1k)-(1m) plan operations within the allowable hydrogen storage range $h^{\rm{tank}}_{kt}$ of the tank with electrolyzer efficiency $\eta^{\rm{el}}$. 
Here, $M^{\rm{H_{2}}}$ represents the heat value of hydrogen, indicating the energy released upon hydrogen combustion and quantifying the conversion between hydrogen and electric power \cite{P2G_7}.

3) VPP Bidding Constraints: Acting as a single generator, the VPP determines its bid quantity for the market by aggregating the bid schedules of its individual resources to optimize its collective output.
\begin{align}\label{eq:VPP Bidding}
\overline{g}^{\rm{VPP}}_{uts} = & \sum_{i \in \cal{I}} \hat{p}^{\rm{g}}_{its} \!+\! \sum_{j \in \cal{J}} ( p^{\rm{dis}}_{jt} \!-\! p^{\rm{ch}}_{jt} ) \!+\! \sum_{k \in \cal{K}}( \eta^{\rm{fc}} M^{\rm{H_{2}}} h^{\rm{fc}}_{kt} \!-\!  p^{\rm{el}}_{kt} ), \nonumber \\
&\forall t \in {\cal T}, \forall s \in {\cal S},
\end{align}
Equation (1n) represents the bid quantity of VPP for the market participation as a single generator under each scenario, aggregating PV and WT generation forecasts with the bid scheduling of ESS and HSS.

4) CVaR Constraints: The risk of profit is considered within the problem formulation by incorporating CVaR at the α-confidence level.
The parameters in the objective function (1a) are as follows: $a \in (0, 1)$, representing the confidence level; $\beta \in [0, \infty)$, a risk weighting factor for integrating risk into the objective function (1); $\eta$, the value-at-risk (VaR).
\begin{align}\label{eq:VPP Bidding 2}
&\eta \! - \! \sum_{t \in  {\cal T}} (
\lambda^{\rm{E}}_{b(\mathrm{VPP}) ts}  g^{\rm{VPP}}_{uts} \!+ \! \sum_{i \in {\cal I }} \kappa^{\mathrm{REC}}_{ts} p^{\rm{g}}_{its} ) \!=\! \epsilon_{s}, \forall s \in {\cal S},   \\
& \epsilon_{s} \geq 0, \quad \forall s \in {\cal S}.
\end{align}
\end{subequations}
Equations (1o) and (1p) enable the linear formulation of CVaR by defining the nonnegative auxiliary variable $\epsilon_{s}$, which captures deviations from the VaR to ensure efficient CVaR quantification and support risk assessment.

\subsection{Lower-level Problem: Market Clearing Model}
In the lower-level problem, the strategic VPP and other non-strategic participants engage in the market to determine generation quantities and prices. 
\begin{subequations}
\begin{align}\label{eq:LL}
\max_{\Xi^{\rm{LL}}} \quad & \! \sum_{u \in \cal{U}} \sum_{t \in \cal{T}} \sum_{s \in \cal{S}}  -C_{u}g_{uts} \\
\text{s.t.} \quad &\! (\xi_{lts}) \!:\! f_{lts} \!=\! \frac{1}{X_{l}}(\theta_{o(l)} - \theta_{e(l)}), \forall  l \in {\cal L}, \forall t \in {\cal T}, \forall s \in {\cal S}, \! \\
&\! (\lambda^{\mathrm{E}}_{bts}) :  \sum_{u \in {\cal U}_{b}}{g_{uts}} \!-\! \sum_{l \vert o(l) = b} f_{lts} \!+\! \sum_{l \vert e(l) = b} f_{lts} \!-\! D_{bts}  = 0, \nonumber \\
&\hspace{3cm} \forall b \in {\cal B},  \forall t \in {\cal T}, \forall s \in {\cal S},  \\
&\! (\underline{\delta}_{lts}, \overline{\delta}_{lts}) : \underline{F_{l}} \leq f_{lts} \leq \overline{F_{l}},
\quad \forall l \in {\cal L}, \forall t \in {\cal T}, \forall s \in {\cal S}, \\
&\! (\underline{\gamma}_{uts}, \overline{\gamma}_{uts}) : 0 \leq g_{uts} \leq \overline{g}_{uts},  \forall u \in {\cal U}, \forall t \in {\cal T}, \forall s \in {\cal S}, \!\!
\end{align}
\end{subequations}
where  $\Xi^{\rm{LL}} \coloneqq \{g_{uts}, \theta_{bts}, f_{lts} \}$.
The objective function (2a) maximizes social welfare, which minimizes operational costs.
Equation (2b) describes the DC power flow model, while the power balance across the network is maintained in (2c), where $o(l)$ and $e(l)$ denote originating and ending nodes of line $l$, respectively.
The linearity of DC power flow model allows the application of Karush-Kuhn-Tucker (KKT) conditions, where strong duality secures a global optimum for the lower-level problem.
Equations (2d) and (2e) define constraints on generation outputs and line capacities to ensure operational reliability.
The dual variables in (2b)-(2e) are specified in parentheses preceding each equation.
The dual variable $\lambda^{\mathrm{E}}_{bts}$ in (2c) is the locational marginal price in the market.

\subsection{Equivalent Bi-Level Problem}
Since the lower-level problem in (2a)-(2e) is linear, then equation (2) can be equivalently reformulated as a mathematical program with equilibrium constraints (MPEC) using the lower-level KKT optimality conditions as followed \cite{VPP_7, VPP_13}.
\begin{subequations}
\begin{align}\label{eq:Bi-level}
\max_{\Xi^{\rm{BL}}} \quad & 
    \sum_{s \in {\cal S}}  \sum_{t \in  {\cal T}} \omega_{s}  (
    \lambda^{\rm{E}}_{b(\mathrm{VPP})ts} g^{\rm{VPP}}_{uts} 
    + \sum_{i \in {\cal I }} \kappa^{\mathrm{REC}}_{ts} p^{\rm{g}}_{its}  ) \nonumber \\
    & + \beta \cdot ( \eta - \frac{1}{1-\alpha}\sum_{s \in {\cal S}  } \omega_{s} \epsilon_{s}  )
\end{align}
\vspace{-4mm}
\begin{align}
&\text{subject to:} \nonumber\\
&\text{Upper-level Constraints:} \,\, \text{Equations (1b)-(1p)} \\
&\text{Lower-level Equality Constraints:} \,\, \text{Equations (2b)-(2c)} \\
&\text{Lower-level KKT Conditions:} \nonumber \\
& \xi_{lts} - \lambda^{\mathrm{E}}_{o(l) = bts} + \lambda^{\mathrm{E}}_{e(l) = bts} + \underline{\delta}_{lts} - \overline{\delta}_{lts} = 0, \nonumber \\
&\hspace{4cm} \forall l \in {\cal L}, t \in {\cal T}, \forall s \in {\cal S}, \\
& -C_{u} + \lambda^{\mathrm{E}}_{b(u)ts}  + \underline{\gamma}_{uts} - \overline{\gamma}_{uts} = 0, \nonumber \\
&\hspace{4cm} \forall u \in {\cal U}, t \in {\cal T}, \forall s \in {\cal S}, \\
& \sum_{l \vert o(l) = b}  ( -\frac{ \xi_{lts} } {X_{l}}  ) \!+\! \sum_{l \vert e(l) = b} ( \frac{ \xi_{lts} } {X_{l}} ) \!=\! 0, \forall b \in {\cal B}, t \in {\cal T}, \forall s \in {\cal S},\hfill \\
& 0 \leq f_{lts} - \underline{F_{l}} \perp \underline{\delta}_{lts} \geq 0,
\quad \forall l \in {\cal L}, t \in {\cal T}, \forall s \in {\cal S},  \\
& 0 \leq -f_{lts} + \overline{F_{l}} \perp \overline{\delta}_{lts} \geq 0,
\quad \forall l \in {\cal L}, t \in {\cal T}, \forall s \in {\cal S},  \\
& 0 \leq g_{uts} - 0 \perp \underline{\gamma}_{uts} \geq 0,
\quad \forall u \in {\cal U} , t \in {\cal T}, \forall s \in {\cal S},  \\
& 0 \!\leq\! \!- g_{uts} + \overline{g}_{uts} \perp \overline{\gamma}_{kts} \geq 0, \!\!\!\!
\quad \forall u \in {\cal U}, t \in {\cal T}, \forall s \in {\cal S}, \!\!\!
\end{align}
\end{subequations}
where \!\! $\Xi^{\rm{BL}} \!\!\coloneqq\!\!  \{  \xi_{lts}, \lambda^{\mathrm{E}}_{bts} \!\in\! \mathbb{R}; \, \underline{\delta}_{lts}, \overline{\delta}_{lts}, \underline{\gamma}_{uts}, \overline{\gamma}_{uts} \!\!\geq\!\! 0 \} \!\cup \Xi^{\rm{UL}} \!\cup\! \Xi^{\rm{LL}}. $
The stationarity conditions are specified in (3d)–(3f), and equations (3g)-(3j) denote the complementary slackness of the lower-level inequality constraints. Here, $\perp$ in (3g)–(3j) denotes orthogonality, meaning that $x \perp y$ implies $x^{\top} y = 0$.

\section{Case Study}\label{Sec:Case}
\begin{figure}
	\centering
    \vspace{-3mm}
\includegraphics[width=1\columnwidth]{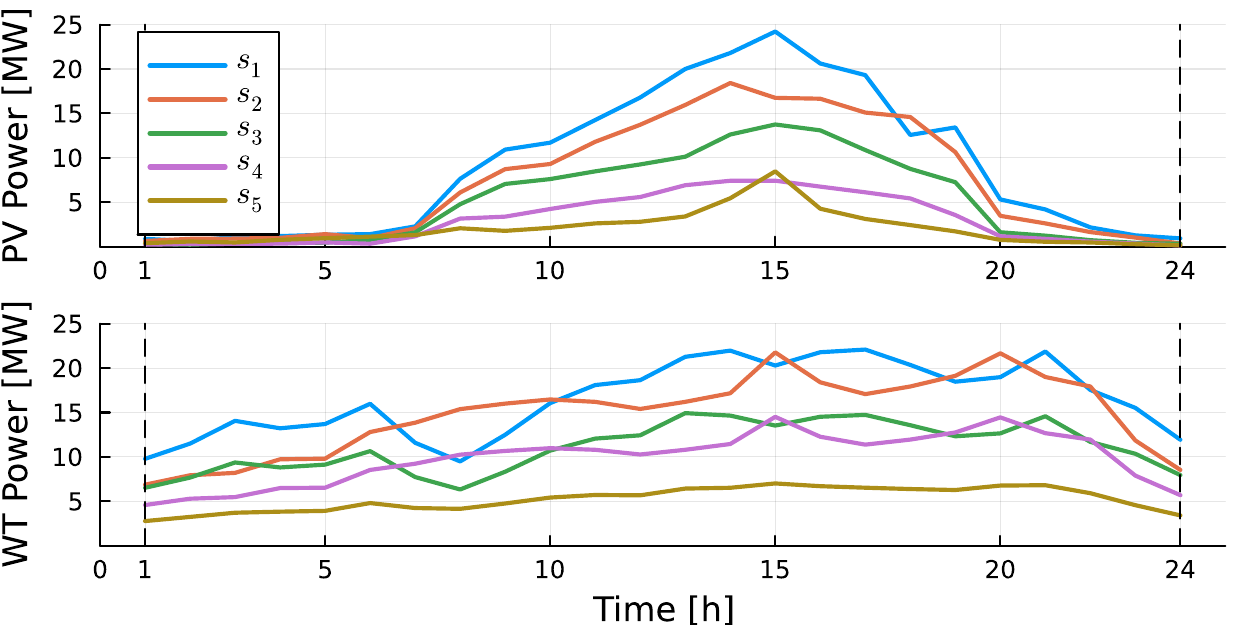}
\vspace{-4mm}
	\caption{\small The generation forecasts of PV, WT for each scenario $s$. \\ The probability of scenario $\omega_{s=1,\dots,5} = \{0.3, 0.25, 0.2, 0.15, 0.1 \}.$
\vspace{-4mm}}
	\label{fig:Forecasts}
\end{figure}

\vspace{-2mm}
The optimal bidding strategy was evaluated on a VPP portfolio comprising PV, WT, ESS, and HSS, with each resource’s profile summarized in Table ~\ref{table_1}.
To compare the efficacy of ESS and HSS, identical power limits for absorption and consumption were set, assuming that the LCOS of HSS is comparable to that of ESS.
For instance, power consumption for ESS charging and HSS electrolyzer operation is both set at 10 MW, while hydrogen consumption in the HSS fuel cell is limited to about 400 kg to match ESS discharging output 10 MW.
Figure ~\ref{fig:Forecasts} represents the scenario-based generation forecasts for VPP resources PV and WT, with the probabilities $\omega_{s}$ for each of the five scenarios.
And the modified IEEE 14-bus system is employed to verify the proposed VPP bidding strategy, ensuring a realistic representation of VPP operation.

\begin{table}[t]
\centering
\captionsetup{justification=centering, labelsep=period, font=footnotesize, textfont=sc}
\caption{VPP DER Profile}
\setlength{\tabcolsep}{13pt}
\begin{tabular}{l l}
\toprule
\textbf{DER} & \textbf{Value} \\ 
\midrule
PV, WT capacity & 20 {[}MW{]}\\ 
ESS discharging,charging limit $\overline{p}^{\rm{dis}}_{jt}, \overline{p}^{\rm{ch}}_{jt} $ & 10 {[}MW{]}\\ 
ESS maximum storage  $\overline{e}_{j}$ & 40 {[}MWh{]}\\
ESS efficiency $\eta^{\rm{dis}}_{j}$, $\eta^{\rm{ch}}_{j}$  &80 {[}\%{]}\\ 
HSS electrolyzer limit $\overline{p}^{\rm{el}}_{kt}, $ & 10 {[}MW{]}\\
HSS fuel cell limit $\overline{h}^{\rm{fc}}_{kt}, $ & 400 {[}kg{]}\\ 
HSS maximum storage  $\overline{h}_{k}$ & 2 {[}ton{]} \\
HSS efficiency $\eta^{\rm{el}}_{k}$, $\eta^{\rm{fc}}_{k}$  & 70, 60 {[}\%{]}\\ 
Heat value of hydrogen $M^{\rm{H_{2}}}$ & 0.033 {[}MWh/kg{]}\\
\bottomrule
\end{tabular}
\label{table_1}
\end{table}

\vspace{-2mm}
\subsection{Impact of Portfolio Configurations on Profitability for VPP: Highlighting Hydrogen Storage System}

\begin{table}[b]
\vspace{-1mm}
\captionsetup{justification=centering, labelsep=period, font=footnotesize, textfont=sc}
\centering
\caption{VPP DER Portfolio and Expected Revenue} 
\setlength{\tabcolsep}{18pt}
\begin{tabular}{c c @{\hskip 30 pt} c}
\toprule
\textbf{Case}& \textbf{DER portfolio} & \textbf{Expected revenue {[}$\$${]}} \\ 
\midrule
(1) & PV + WT & 3,710 ($-$) \\ 
(2) & PV + WT + ESS  & 4,405 (18.7$\%\uparrow$) \\ 
(3) & PV + WT + HSS  & 4,607 (24.2$\%\uparrow$) \\ 
\bottomrule
\end{tabular}
\label{table_2}
\end{table}

\begin{figure}[t]
	\vspace{-3mm}
	\centering
    \captionsetup{justification=centerlast}
	\includegraphics[width=1\columnwidth]{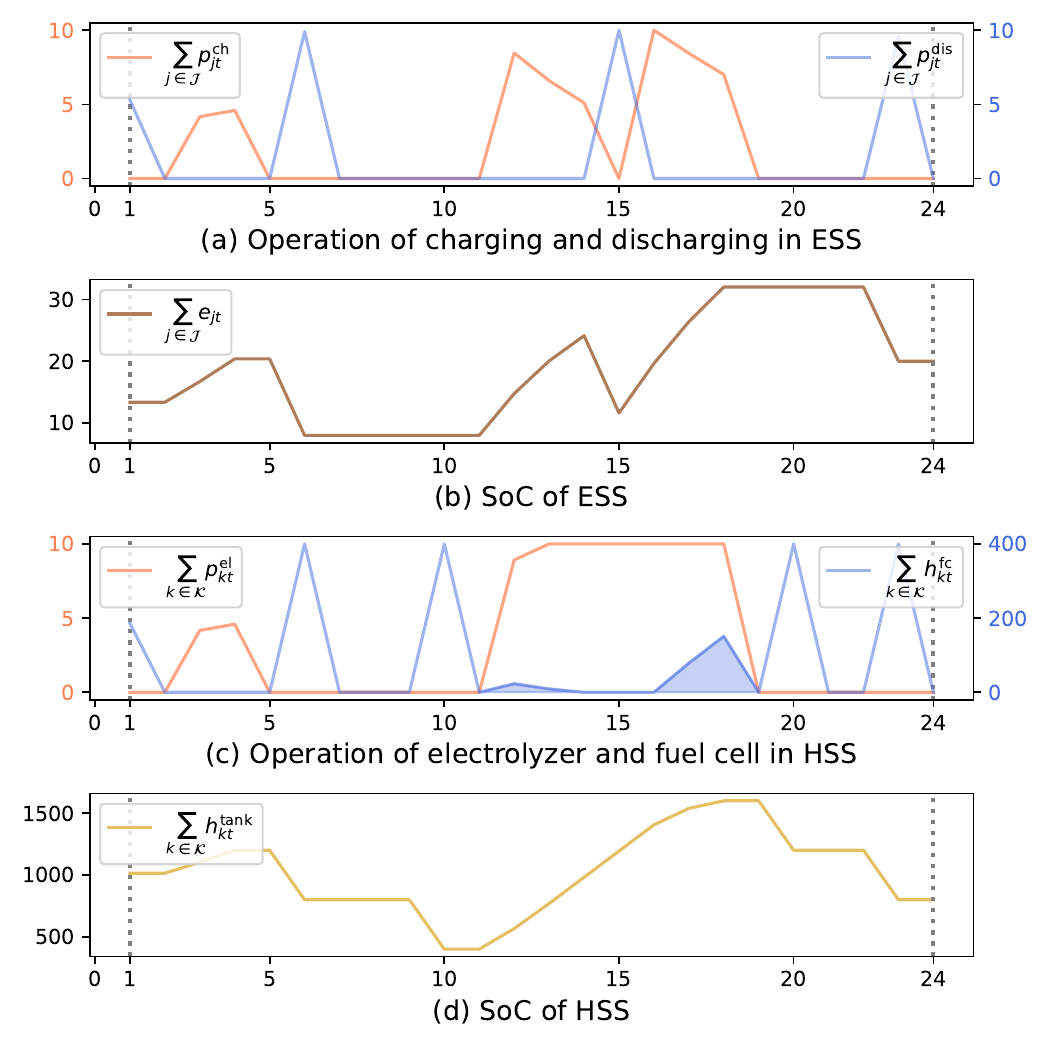}
	\caption{\small Comparison of ESS and HSS operation.
    (a) Operation of charging and discharging in ESS
    (b) SoC of ESS
    (c) Operation of electrolyzer and fuel cell in HSS
    (d) SoC of HSS. \vspace{-5mm}}
	\label{fig:result_1}
\end{figure}

To investigate the impact of VPP resource configurations on revenue, the expected revenue is analyzed under three cases:
(1) with PV and WT only, (2) inclusion of ESS, and (3) inclusion of HSS. Here, $\beta$ is set to 0, indicating a risk-neutral approach in the bidding strategy.
As shown in Table ~\ref{table_2}, the configuration that includes only PV and WT (Case 1) leads to the lowest expected profit, due to the limited ability to capitalize on market price variations, which is crucial for maximizing revenue.
However, the integration of ESS and HSS enhances the expected revenue, primarily due to the ability of these storage systems to support effective arbitrage strategies that respond dynamically to changes in market prices (Cases 2 and 3).
Moreover, substituting ESS with an equivalent capacity of HSS within the VPP, leads to a further increase in the overall revenue, 24\% increase in this case.
The results suggest that, when considering energy storage options within VPP, installing HSS—given its expected lower LCOS compared to ESS by 2030—provides a strategic advantage for enhancing profitability.
Figure ~\ref{fig:result_1} compares the operational characteristics of ESS and HSS in power delivery.
The scheduling of the electrolyzer and fuel cell within the HSS, shown in Figure \ref{fig:result_1}(c), demonstrates concurrent operation in the blue-shaded area, whereas, as depicted in Figure \ref{fig:result_1}(a), such concurrent operation is not feasible for ESS.
With round-trip capability, HSS supports a more sophisticated bidding strategy than ESS. Its electrolyzer enables continuous charging while the fuel cell operates concurrently, leading to improved profitability.

\begin{table}[b]
\vspace{-1mm}
\captionsetup{justification=centering, labelsep=period, font=footnotesize, textfont=sc}
\centering
\caption{Size of HSS tank and Expected Revenue} 
\setlength{\tabcolsep}{18pt}
\begin{tabular}{c c @{\hskip 30 pt} c}
\toprule
\textbf{Case}& \textbf{Storage size} & \textbf{Expected revenue {[}$\$${]}} \\ 
\midrule
(4) & $\overline{h_{k}}$ & 4,607 ($-$) \\ 
(5) & 0.5 $\overline{h_{k}}$  & 4,348 (5.6$\%\downarrow$) \\ 
(6) & 2 $\overline{h_{k}}$  & 4,833 (4.9$\%\uparrow$) \\ 
\bottomrule
\end{tabular}
\label{table_3}
\end{table}

\vspace{-2mm}
\subsection{Effect of the Risk Aversion Factor $\beta$}

\begin{figure}[t]
	\centering
    \captionsetup{justification=centerlast}
	\includegraphics[width=1\columnwidth]{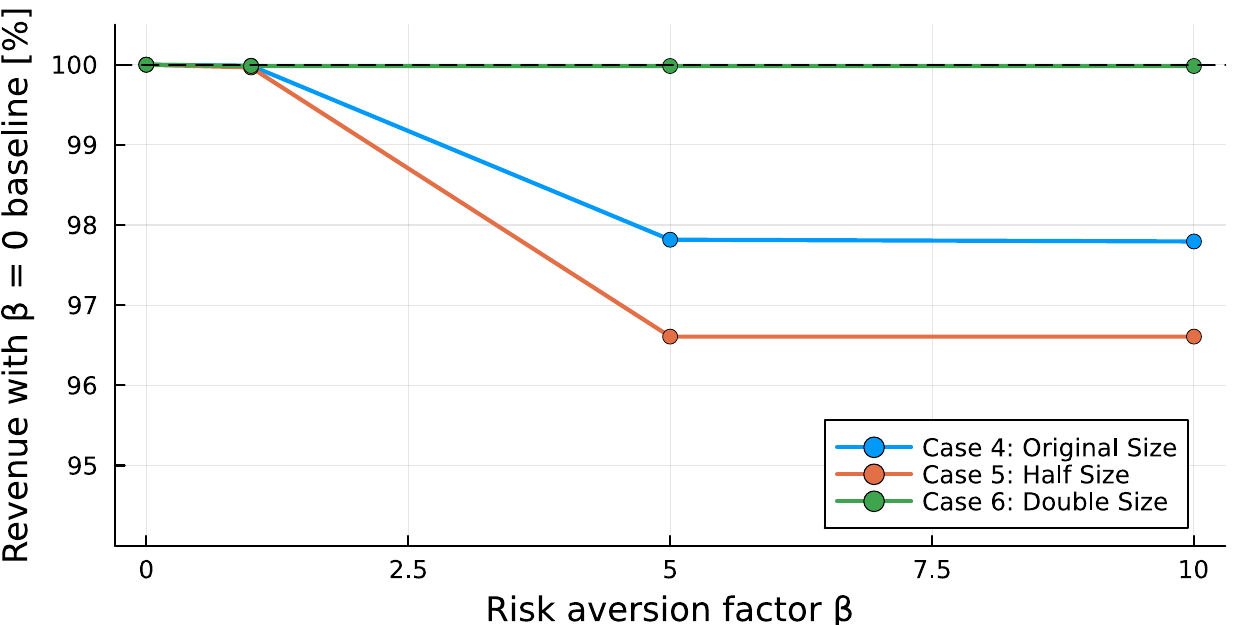}
	\caption{ Effect of $\beta$ on revenue across different storage sizes.\vspace{-7mm}}
	\label{fig:result_2}
\end{figure}

\begin{figure}[b]
	\centering
    \captionsetup{justification=centerlast}
    \vspace{-7mm}
	\includegraphics[width=1\columnwidth]{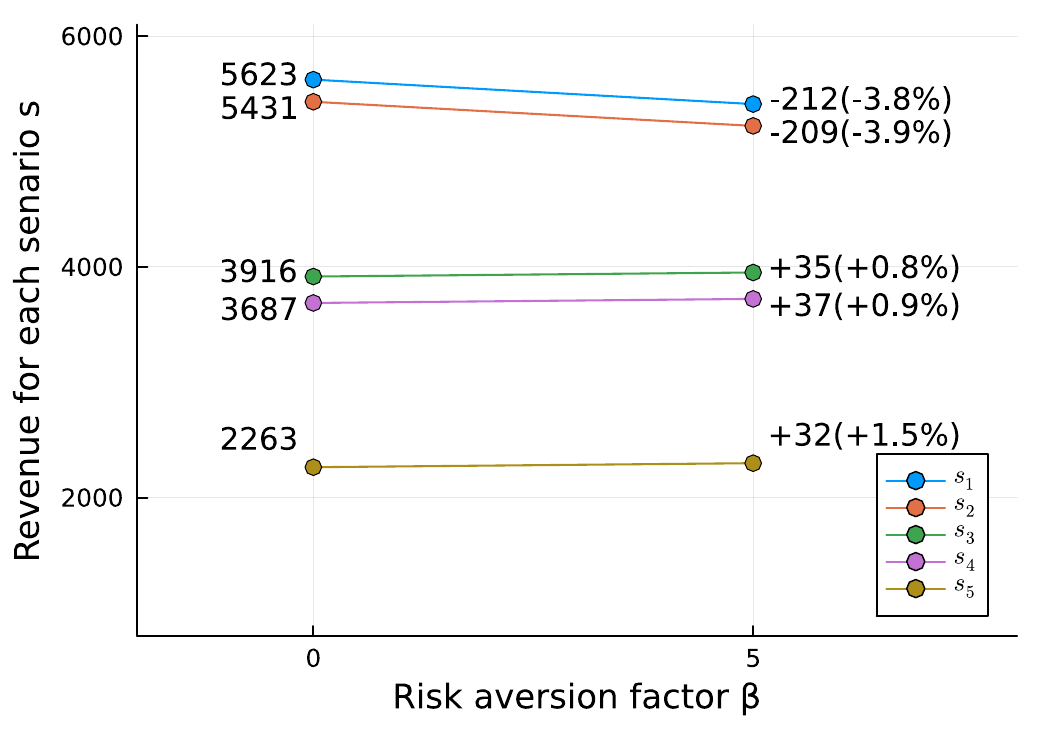}
	\caption{ Effect of $\beta$ on revenue for different realized scenarios.}
	\label{fig:result_3}
\end{figure}

To analyze the profitability of a bidding strategy considering risk aversion,
the revenue of VPP was analyzed across different storage size of tank.
The portfolio of VPP consists of PV, WT, and HSS as in case (3) which maintains an identical power capacity for both the electrolyzer and the fuel cell, yet features a different storage size.
Table ~\ref{table_3} illustrates revenue across different storage sizes of hydrogen tank for a risk-neutral bidding strategy, with $\beta = 0$.
A comparison of expected revenue across different storage sizes suggests that, when accounting for installation costs, smaller storage seems to offer a more favorable economic outcome.
However, Figure ~\ref{fig:result_2} illustrates the comparison of revenue changes across different storage sizes as a function of $\beta$, based on the expected revenue in Table ~\ref{table_3}.
Increasing $\beta$ leads to a general decline in expected revenue, reflecting heightened risk aversion.
A comparison of expected profit for different storage sizes shows that with increasing  $\beta$, the higher storage size (Case 6) exhibits lower sensitivity in expected revenue under risk aversion than lower storage size (Case 4 and 5).
The findings suggest that energy storage not only increases revenue but also improves robustness against profitability through risk aversion strategy.

Figure ~\ref{fig:result_3} illustrates the comparison of revenue for each realized scenarios as a function of the risk aversion factor $\beta$ in Case 4.
As the subscript for scenario s increases from 1 to 5, PV and WT generation decrease, as shown in Figure ~\ref{fig:Forecasts}, resulting in a corresponding decline in revenue across scenarios.
For scenarios $s_{1}$ and $s_{2}$, high PV and WT generation encourage aggressive market strategies, causing revenue to decline with higher $\beta$.
The revenue in scenarios  $s_{3}$ and $s_{4}$ exhibits minimal sensitivity to variations in revenue.
Conversely, in the lowest profit scenario $s_5$, where PV and WT generation are low, higher $\beta$ rather increase the gained revenue, suggesting a beneficial effect of risk aversion.
These results suggest that establishing a risk-aversion strategy aligned with market profitability conditions is beneficial for maximizing the revenue.

\section{Conclusion}\label{Sec:Conclusion}
\vspace{-2mm}
This paper presents a risk-aware bi-level bidding strategy for VPP with a P2H system, incorporating CVaR to balance profit maximization with risk mitigation.
The upper-level problem optimizes the revenue, while the lower-level ensures feasibility under market-clearing constraints and CVaR-based risk aversion improves financial hedging effectiveness. Case study results illustrate that integrating a P2H system enhances the revenue, while providing robustness in risk hedging. And selecting an appropriate risk aversion factor further optimizes revenue by aligning VPP bidding with market profitability. Future research could incorporate real-time market dynamics to help VPP adapt to RES fluctuations, enhancing efficiency, while deploying with hydrogen markets may unlock hydrogen’s role as a flexible energy carrier.
\vspace{-2mm}

\balance
\bibliographystyle{IEEEtran}
\bibliography{ref.bib}

\begin{thebibliography}{10}
\providecommand{\url}[1]{#1}
\csname url@samestyle\endcsname
\providecommand{\newblock}{\relax}
\providecommand{\bibinfo}[2]{#2}
\providecommand{\BIBentrySTDinterwordspacing}{\spaceskip=0pt\relax}
\providecommand{\BIBentryALTinterwordstretchfactor}{4}
\providecommand{\BIBentryALTinterwordspacing}{\spaceskip=\fontdimen2\font plus
\BIBentryALTinterwordstretchfactor\fontdimen3\font minus \fontdimen4\font\relax}
\providecommand{\BIBforeignlanguage}[2]{{%
\expandafter\ifx\csname l@#1\endcsname\relax
\typeout{** WARNING: IEEEtran.bst: No hyphenation pattern has been}%
\typeout{** loaded for the language `#1'. Using the pattern for}%
\typeout{** the default language instead.}%
\else
\language=\csname l@#1\endcsname
\fi
#2}}
\providecommand{\BIBdecl}{\relax}
\BIBdecl

\bibitem{VPP_9}
J.~Downing, N.~Johnson, M.~McNicholas, D.~Nemtzow, R.~Oueid, J.~Paladino, and E.~B. Wolfe, ``Pathways to commercial liftoff: Virtual power plants,'' \emph{US Department of Energy Report}, 2023.

\bibitem{VPP_2}
Q.~Wu and C.~Li, ``A bi-level optimization framework for the power-side virtual power plant participating in day-ahead wholesale market as a price-maker considering uncertainty,'' \emph{Energy}, vol. 304, p. 132050, 2024.

\bibitem{VPP_7}
E.~G. Kardakos, C.~K. Simoglou, and A.~G. Bakirtzis, ``Optimal offering strategy of a virtual power plant: A stochastic bi-level approach,'' \emph{IEEE Transactions on Smart Grid}, vol.~7, no.~2, pp. 794--806, 2015.

\bibitem{VPP_10}
N.~Pourghaderi, M.~Fotuhi-Firuzabad, M.~Kabirifar, M.~Moeini-Aghtaie, M.~Lehtonen, and F.~Wang, ``Reliability-based optimal bidding strategy of a technical virtual power plant,'' \emph{IEEE Systems journal}, vol.~16, no.~1, pp. 1080--1091, 2021.

\bibitem{VPP_15}
E.~Mashhour and S.~M. Moghaddas-Tafreshi, ``Bidding strategy of virtual power plant for participating in energy and spinning reserve markets—part i: Problem formulation,'' \emph{IEEE Transactions on Power Systems}, vol.~26, no.~2, pp. 949--956, 2010.

\bibitem{VPP_12}
S.~R. Dabbagh and M.~K. Sheikh-El-Eslami, ``Risk assessment of virtual power plants offering in energy and reserve markets,'' \emph{IEEE Transactions on Power Systems}, vol.~31, no.~5, pp. 3572--3582, 2015.

\bibitem{Book_1}
A.~J. Conejo, M.~Carri{\'o}n, J.~M. Morales \emph{et~al.}, \emph{Decision making under uncertainty in electricity markets}.\hskip 1em plus 0.5em minus 0.4em\relax Springer, 2010, vol.~1.

\bibitem{P2G_1}
{Department of Energy, U.S.}, ``{U.S. National clean hydrogen strategy and roadmap},'' Technical Report, Tech. Rep., 2022.

\bibitem{P2G_2}
P.~J. Balducci, D.~Wu, T.~Ramachandran, A.~M. Campbell, V.~Fotedar, K.~Mongird, S.~Huang, D.~Bhatnagar, S.~Jones, C.~Smith \emph{et~al.}, ``Power-to-gas system valuation. final report,'' Pacific Northwest National Lab.(PNNL), Richland, WA (United States), Tech. Rep., 2020.

\bibitem{P2G_10}
V.~Viswanathan, K.~Mongird, R.~Franks, X.~Li, V.~Sprenkle, and R.~Baxter, ``2022 grid energy storage technology cost and performance assessment,'' \emph{Energy}, vol. 2022, 2022.

\bibitem{P2G_9}
V.~J{\"u}lch, ``Comparison of electricity storage options using levelized cost of storage (lcos) method,'' \emph{Applied energy}, vol. 183, pp. 1594--1606, 2016.

\bibitem{P2G_3}
Y.~Meng, S.~Fan, X.~Zheng, J.~Xiao, H.~Zhou, and G.~He, ``Optimal operation of virtual power plant considering power-to-hydrogen systems,'' in \emph{2022 IEEE power \& energy society general meeting (PESGM)}.\hskip 1em plus 0.5em minus 0.4em\relax IEEE, 2022, pp. 1--5.

\bibitem{P2G_4}
Q.~Wu and C.~Li, ``A bi-level optimization framework for the power-side virtual power plant participating in day-ahead wholesale market as a price-maker considering uncertainty,'' \emph{Energy}, vol. 304, p. 132050, 2024.

\bibitem{P2G_6}
L.~Rodrigues, T.~Soares, I.~Rezende, J.~Fontoura, and V.~Miranda, ``Virtual power plant optimal dispatch considering power-to-hydrogen systems,'' \emph{International Journal of Hydrogen Energy}, vol.~68, pp. 1019--1032, 2024.

\bibitem{P2G_7}
F.~Dawood, M.~Anda, and G.~Shafiullah, ``Hydrogen production for energy: An overview,'' \emph{International Journal of Hydrogen Energy}, vol.~45, no.~7, pp. 3847--3869, 2020.

\bibitem{VPP_13}
J.~Kim, R.~Mieth, and Y.~Dvorkin, ``Computing a strategic decarbonization pathway: A chance-constrained equilibrium problem,'' \emph{IEEE Transactions on Power Systems}, vol.~36, no.~3, pp. 1910--1921, 2020.

\end{thebibliography}

\end{document}